\documentclass
[aps,prl,twocolumn,floatfix,english,showpacs,10pt,superscriptaddress]{revtex4-2}%
\usepackage{graphicx}
\usepackage{amsmath}
\usepackage{physics}
\usepackage{amssymb}
\usepackage{colordvi}
\usepackage{verbatim}
\usepackage{xcolor}
\usepackage{mathrsfs}
\usepackage{epsfig}
\usepackage{lipsum}
\usepackage{amsfonts}
\usepackage[unicode=true, breaklinks=false, pdfborder={0 0 1}, backref=false,
colorlinks=true, linkcolor=blue, urlcolor=blue, citecolor=blue]{hyperref}%
\providecommand{\U}[1]{\protect\rule{.1in}{.1in}}
\setcitestyle{numbers,square}
\begin{document}

\title{Efficient Gating of Magnons by Proximity Superconductors}
\author{Tao Yu}
\email{taoyuphy@hust.edu.cn}
\affiliation{School of Physics, Huazhong University of Science and Technology, Wuhan 430074, China}
\author{Gerrit E. W. Bauer}
\email{G.E.W.Bauer@imr.tohoku.ac.jp}
\affiliation{WPI-AIMR and Institute for Materials Research and CSRN, Tohoku
	University, Sendai 980-8577, Japan}
\affiliation{Zernike Institute for Advanced Materials, University of Groningen, 9747 AG Groningen, Netherlands}
\affiliation{Kavli Institute for Theoretical Sciences, University of the Chinese Academy of Sciences, Beijing 100190, China}

\date{\today }

\begin{abstract}
Electrostatic gating confines and controls the transport of electrons in
integrated circuits. Magnons, the quanta of spin waves of the magnetic order,
are promising alternative information carriers, but difficult to gate. Here we
report that superconducting strips on top of thin magnetic films can totally
reflect magnons by its diamagnetic response to the magnon stray fields. The
induced large frequency shifts unidirectionally block the magnons propagating
normal to the magnetization. Two superconducting gates parallel to the
magnetization create a magnonic cavity. The option to gate coherent magnons
adds functionalities to magnonic devices, such as reprogrammable logical
devices and increased couplings to other degrees of freedom.

\end{abstract}
\maketitle

\textit{Introduction.}---Electrostatic gating provides a tunable potential to
control the transport channels of electrons in field effect transistors
\cite{mesoscopic_physics}. Potential barriers size-quantize the electronic
wave function, leading to quantum point-contacts, wires, and dots with
application in quantum technology
\cite{quantum_dots,quantum_dots_qubit_1,quantum_dots_qubit_2}.
\textquotedblleft Magnonics\textquotedblright\ employs the magnon
quasiparticles, i.e., the bosonic quanta of the spin wave excitations of the
magnetic order, as information carriers in a low-power alternative to
conventional electronics
\cite{magnonics_1,magnonics_2,magnonics_3,magnonics_4,magnonics_5,magnonics_roadmap,Walker,DE,new_book,cavity_magnonics}%
. However, magnons cannot be gated, blocked, or trapped as easily as
electrons, so many mature concepts of electronics cannot directly be applied
in magnonics. Chumak \textit{et al}. \cite{Chumak_transistor} propose an
alternatives to gating in a magnon transistor by inserting a magnonic crystal
into a film of yttrium iron garnet (YIG) that controls the transmission of
coherent magnons. Current-biased heavy metal contacts can modulate the
incoherent magnon currents in a YIG channel by spin injection or heating
\cite{Cornelissen_gate,Munich_gate,Liu_gate}. However, both mechanisms do not
create potential barriers, the modulation efficiency is low, and power demands
are high. Electrostatic gates on magnetic semiconductors can locally suppress
magnetism \cite{Ohno_1,Xiaodong_Xu}, but at present the magnetic quality of
available materials remains wanting.

The dispersion of long-wavelength magnetostatic spin waves in thick films or
magnonic crystals can be modulated in a non-reciprocal fashion by the
electromagnetic interaction with metallic gates. The latter can be modelled as
thick perfect \cite{perfect_conductor_1} or Ohmic conductors
\cite{conductor_5,conductor_1,conductor_2,conductor_3,conductor_4}.
Superconducting gate can have profound effects as well
\cite{magnon_superconductor_1,magnon_superconductor_2,magnon_superconductor_3,magnon_superconductor_4}%
. Recent research on ultrathin films accumulates evidence that the physics
changes drastically when thicknesses are reduced down to nanometer scale. In
this limit surface and volume modes merge into perpendicular standing spin
waves (PSSW) with spectra dominated by the exchange interaction. The GHz
dynamics is confined to the lowest PSSW with nearly uniform amplitude normal
to the film \cite{subbands_1,subbands_2,Chiral_pumping_Yu,Haiming_exp_grating}%
. In contrast to conventional wisdom for thick films, a metallic cap atop
ultrathin films of YIG only enhances the damping of spin waves
\cite{Bertelli,damping_PRApplied} with minor effects on their dispersion. The
magnon conductivity in high-quality nanometer films \cite{Xiang_Yang} reaches
record values because of the onset of two-dimensional diffuse magnon
transport. Moreover, stray fields emitted by spin waves in ultrathin films
have a relatively short wave length and thereby cannot penetrate deeply into
metallic gates, a limit not accounted for by previous theories.

\begin{figure}[th]
{\normalsize \begin{centering}
			\includegraphics[width=8.6cm]{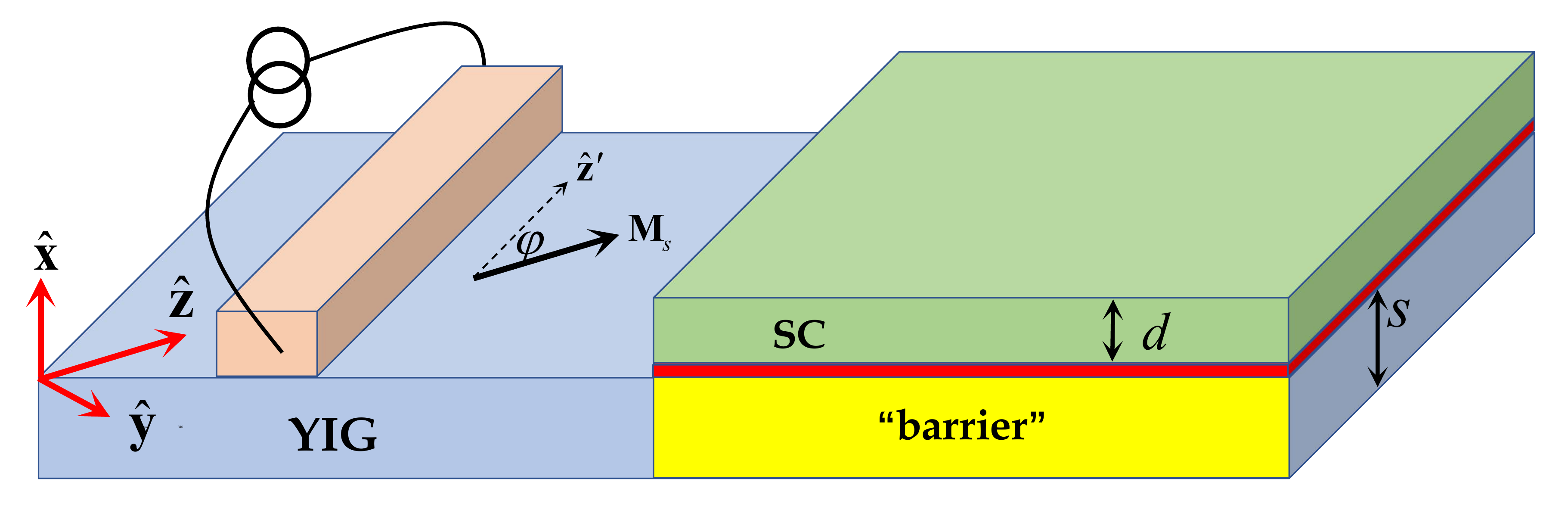}
			\par\end{centering}
}\caption{Unidirectional blocking of magnons by a superconducting
metal gate on a thin magnetic insulator that provides an effective
\textquotedblleft barrier\textquotedblright. An inserted thin insulating layer
between the conductors and magnetic films (red) can strongly suppress the
interfacial exchange interaction. The stripline on the left acts
as injector and detector of coherent magnons. We address reflection as a
function of angle of an in-plane magnetic field that rotates the magnetization
$\mathbf{M}_{s}$.}%
\label{model}%
\end{figure}

In this Letter, we report that floating superconducting films on top of
ultrathin films of a magnetic insulator such as YIG, as illustrated in
Fig.~\ref{model}, induce chiral frequency shifts of tens of GHz that
correspond to magnetic fields of $O(0.1\,\mathrm{T})$. The chirality is caused
by constructive interference of the Oersted fields of the spin-wave dipolar
and induced eddy currents in the superconductors. The gate generates an
effective barrier that is dynamic, i.e., it depends on the magnon frequency
and propagation direction. A wide superconducting gate totally reflects
coherent magnons in the microwave band propagating normal to the
\textquotedblleft up\textquotedblright\ magnetization, but
transmits them when magnetization is flipped. We can thereby control the
magnon current by either modulating the superconductivity in the gates or
rotating the equilibrium magnetization. The on-off ratio is nearly unity over
tens of GHz, in contrast to narrow-band resonant coupling effects
\cite{magnon_trap,spin_wave_diode_1}. Our set-up is exceedingly simple and
does not require spin-orbit interactions, in contrast to metal-based designs
\cite{diode_PRX,diode_PRApplied}. Moreover, we predict that two
superconducting strips can nearly perfectly trap spin waves in a
sub-micrometer region forming a magnon waveguide without having to etch the
magnetic film. On-chip implementation of these devices allow to implement
magnonic functionalities such as non-volatility and chirality into quantum technologies.

\textit{Inductive interaction between magnons and superconductors.}---We
consider a floating superconducting (SC) or normal metal (NM) gate with
thickness $d$ on a thin YIG film of thickness $s$ as illustrated in
Fig.~\ref{model}. A thin insulating spacer between them suppresses the
proximity effect by an interfacial exchange interaction. We formulate the
inductive coupling of propagating magnons under the gate that gives rise to
\textquotedblleft eddy current\textquotedblright%
\ \cite{Schoen,damping_PRApplied,Bertelli}. The magnetization is parallel to
and rotates with a weak in-plane applied magnetic field $\mathbf{H}%
_{\mathrm{app}}$, which does not affect the superconductivity. We use a
coordinate system in which $\mathbf{H}_{\mathrm{app}}=H_{\mathrm{app}}%
\hat{\mathbf{z}}$ and an angle $\varphi$ relative to the contacts, while the
film normal is along $\hat{\mathbf{x}}$. We assume that the magnetization
dynamics is dominated by isotropic exchange magnons in the thin films and
disregard the dipolar interaction on the dispersion and small mode ellipticity
as justified in the Supplemental Material \cite{supplement}. With free
boundary conditions \cite{subbands_1,subbands_2}, the exchange spin waves with
amplitudes $m_{y}^{(l)}(\mathbf{k})\approx im_{x}^{(l)}(\mathbf{k})$ are
circularly polarized \cite{supplement}, where $l$ is the band indices
\cite{Doppler,Xiang_Yang}.

The dynamic magnetic stray fields emitted by the spin waves induce eddy
currents in conductors that in turn generate Oersted magnetic fields
$\mathbf{H}$ that affect the spin-wave dynamics. We focus on metallic gates
with $d\sim O(10~\mathrm{nm})$ because currents are then constant over the
film thickness: London's penetration depth of SCs $\lambda_{L}=\sqrt
{m_{e}/(\mu_{0}n_{s}e^{2})}\sim O(100~\mathrm{nm})$
\cite{Kittel_Book,Schrieffer}, where $\mu_{0}$ is the vacuum permeability and
$m_{e}$ is the effective mass of electrons with density $n_{s}$. The
\textquotedblleft skin depth\textquotedblright\ $\delta=\sqrt{2/(\mu_{0}%
\omega\sigma_{c})}$ of NMs with conductivity $\sigma_{c}$ for ac magnetic
fields of frequency $\omega$ \cite{Jackson} is also much larger than $d$ for
good metals at frequencies up to several terahertz (THz). A pioneering study
\cite{perfect_conductor_1} considers \textquotedblleft perfectly
conducting\textquotedblright\ gates with $\sigma_{c}\rightarrow\infty$
corresponding to $\delta\rightarrow0$ and interface boundary condition
$\mathbf{B}_{\perp}={0}$, which is very different from the superconductors
discussed below.

The dipolar stray fields $\mathbf{H}^{(l)}_{\mathbf{k}}(\mathbf{r},t)$ of
spin-wave eigenmodes with momentum $\mathbf{k}$ and frequency $\omega
_{\mathbf{k}}$ above the film \cite{supplement}, i.e., $\mathbf{r}%
=(x>0,\pmb{\rho}=y\hat{\mathbf{y}}+z\hat{\mathbf{z}}),$ are evanescent as
$\sim e^{-\left\vert \mathbf{k}\right\vert x}$ and nearly uniform across a
gate for wave numbers $\left\vert \mathbf{k}\right\vert \ll1/d$
\cite{subbands_2}. When $\left\vert \mathbf{k}\right\vert \ll1/s$ only the
lowest subband contributes. The components of the magnetic field
$\mathbf{H}_{\mathbf{k}}^{(l=0)}$ are locked with momentum, obeying $k_{z}%
{H}_{y}^{(0)}=k_{y}{H}_{z}^{(0)}$ and $\left\vert \mathbf{k}\right\vert
{H}_{x}^{(0)}=ik_{y}{H}_{y}^{(0)}+ik_{z}{H}_{z}^{(0)}$. The stray field
vanishes when $\mathbf{k}=-|k_{y}|\hat{\mathbf{y}},$ so spin waves that
propagate in the negative $\hat{\mathbf{y}}$-direction cannot interact with a
top gate.

According to Maxwell's equations $\nabla\cdot\mathbf{E}=0$ and $\nabla
\times\mathbf{E} =i\omega\mu_{0}\mathbf{H}$, the magnetic stray field of a
spin wave with wavevector $\mathbf{k}$ and frequency $\omega$ generates an
electromotive force (emf) that has the form $\mathbf{E}\propto e^{-\left\vert
\mathbf{k}\right\vert x+i\mathbf{k}\cdot\pmb{\rho}}$ up to high frequencies
$\sigma_{c}/\varepsilon_{0}\sim10^{6}$~THz for typical metallic conductivities
$\sigma_{c}\sim10^{7}~\mathrm{\Omega}^{-1}\mathrm{m}^{-1}$. The normal
component is immediately screened, so $E_{x}=0$, $E_{y}=-\left(  {i\omega
\mu_{0}}/\left\vert \mathbf{k}\right\vert \right)  {H}_{z}^{(0)}$, and
$E_{z}=\left(  {i\omega\mu_{0}}/\left\vert \mathbf{k}\right\vert \right)
{H}_{y}^{(0)}$. By Ohm's law, the induced electric eddy current in a normal
conductor $J_{x}(x,\pmb{\rho})=0$, $J_{y}(x,\pmb{\rho}) =-\left(  {i\sigma
_{c}\omega\mu_{0}}/\left\vert \mathbf{k}\right\vert \right)  {H}_{z}%
^{(0)}(x,\pmb{\rho})$, and $J_{z}(x,\pmb{\rho}) =\left(  {i\sigma_{c}\omega
\mu_{0}}/\left\vert \mathbf{k}\right\vert \right)  {H}_{y}^{(0)}%
(x,\pmb{\rho})$ is perpendicular to the spin-wave propagation direction since
$\mathbf{k}\cdot\mathbf{J}=0$. Abrikosov vortice lattices of type-II
superconductors are sources of periodic magnetic fields that generate tunable
bandgaps in the magnon spectrum \cite{type_II_1}. We will address such effects
and that of different pairing symmetries in the future.

The eddy currents, in turn, generate Oersted magnetic fields that oppose the
original ones (Lenz effect). The effect on the gate itself may be disregarded
since we consider only films much thinner than London's penetration (SC) and
skin (NM) depths. However, they affect the spin wave dynamics by a field-like
(for SC) or a damping-like (for NM) torque that causes a frequency shift (for
SC) and an additional contribution to the Gilbert damping constant (for NM),
respectively. Here we address the full dynamic response by self-consistently
solving the coupled Maxwell and Landau-Lifshitz-Gilbert (LLG) equations. In
the limit that $\left\vert \mathbf{k}\right\vert \ll1/d,$ the currents are
uniform across the film and we may set $x\rightarrow d/2$. The vector
potential generated by the eddy currents then reads in frequency space
\cite{Jackson} $A_{\xi}(\mathbf{r},\omega)=({\mu_{0}}/{4\pi})\int
d\mathbf{r}^{\prime}{J}_{\xi}\left(  x^{\prime}={d}/{2},\pmb{\rho}^{\prime
},\omega\right)  {e^{i\omega|\mathbf{r}-\mathbf{r}^{\prime}|/c}}%
/{|\mathbf{r}-\mathbf{r}^{\prime}|}$. Using the Weyl identity
\cite{nano_optics}, we obtain for $x<0$
\begin{equation}
\hspace{-0.1cm}A_{\xi}(\mathbf{r},\omega)=({\mu_{0}d}/{2}){J}_{\xi}\left(
x^{\prime}={d}/{2},\pmb{\rho},\omega\right)  {e^{|\mathbf{k}|(x-d/2)}%
}/{|\mathbf{k}|}.
\end{equation}
For the LLG equation we require only the transverse ($x$- and $y$-) components
of the Oersted magnetic field $\tilde{\mathbf{H}}(\mathbf{r})=\nabla
\times\mathbf{A}(\mathbf{r})/\mu_{0}$: $\tilde{H}_{y}(\mathbf{r}%
,\omega)=(ik_{y}/|\mathbf{k}|)\tilde{H}_{x}(\mathbf{r},\omega)$ with
\begin{align}
\tilde{H}_{x}(\mathbf{r},\omega)  &  =ie^{|\mathbf{k}|(x-d)}{d}\sigma
_{c}\omega\mu_{0}({1-e^{-|\mathbf{k}|s}})/(4{|\mathbf{k}|})\nonumber\\
&  \times\left(  m_{x}^{\mathbf{k}}(\pmb{\rho},\omega)-im_{y}^{\mathbf{k}%
}(\pmb{\rho},\omega){k_{y}}/{|\mathbf{k}|}\right)  .
\end{align}

The linearized LLG equation in frequency space poses a self-consistency
problem
\begin{align}
-i\omega m_{x}(\mathbf{k})  &  =-i\omega_{\mathbf{k}}(1-i\alpha_{G}%
)m_{x}(\mathbf{k})+\mu_{0}\gamma M_{s}\tilde{H}_{y}(\mathbf{k},\omega
),\nonumber\\
-i\omega m_{y}(\mathbf{k})  &  =-i\omega_{\mathbf{k}}(1-i\alpha_{G}%
)m_{y}(\mathbf{k})-\mu_{0}\gamma M_{s}\tilde{H}_{x}(\mathbf{k},\omega
),\nonumber
\end{align}
that leads to the modified dispersion relation
\begin{equation}
\tilde{\omega}_{\mathbf{k}}=\frac{\omega_{\mathbf{k}}(1-i\alpha_{G}%
)}{1+i\alpha_{m}(|\mathbf{k}|)\left(  {k_{y}}/{|\mathbf{k}|}+{|k_{y}%
|}/{|\mathbf{k}|}\right)  }, \label{frequency_renormalized}%
\end{equation}
where $\alpha_{G}$ is the intrinsic Gilbert damping coefficient and
\begin{equation}
\alpha_{m}(|\mathbf{k}|)=({d}/{4})e^{|\mathbf{k}|\left(  -\frac{s}%
{2}-d\right)  }(1-e^{-|\mathbf{k}|s})\sigma_{c}\mu_{0}^{2}\gamma
M_{s}/|\mathbf{k}|
\end{equation}
is dimensionless. In the limit that $\left\vert \mathbf{k}\right\vert
\ll\{1/d,1/s\}$, $\alpha_{m}(\left\vert \mathbf{k}\right\vert )\rightarrow
ds\sigma_{c}\mu_{0}^{2}\gamma M_{s}/4$. In a normal metal, $\sigma_{c}$ is
real and $\alpha_{m}>0,$ so only magnons with positive $k_{y}$ suffer from an
additional damping $\tilde{\alpha}(\mathbf{k})=\alpha_{m}\cos\theta
_{\mathbf{k}}\left(  1+\mathrm{sgn}(k_{y})\right)  $, where $\cos
\theta_{\mathbf{k}}=k_{y}/\left\vert \mathbf{k}\right\vert $. For a copper
conductor gate with thickness $d=40$\thinspace nm and conductivity $\sigma
_{c}\approx6\times10^{7}$$\,$$\Omega^{-1}$m$^{-1}$ on top of a $s=20$%
\thinspace nm thin YIG film with $\mu_{0}M_{s}=0.18$\thinspace T and
$\gamma=1.82\times10^{11}$\thinspace s$^{-1}$T$^{-1}$, we find $\alpha
_{m}=4.8\times10^{-4},$ which is of the same order as the intrinsic damping.
Bertelli \textit{et al.} \cite{Bertelli} observed a larger $\alpha_{m}%
\sim10^{-2}$ by nitrogen-vacancy center magnetometry, but for thicker films
and only for positive $\mathbf{k}=|k_{y}|\hat{\mathbf{y}},$ therefore did not
yet resolve the damping chirality. In contrast to thick films
\cite{conductor_2}, the gate hardly affects the spin wave dispersion Eq.
(\ref{frequency_renormalized}).

The chiral damping theory above holds for coherent spin waves with well
defined momentum as excited by narrow striplines, and we may expect similar
effects in the diffuse regime of magnon transport. An asymmetry in the
propagation of carriers into opposite directions has been reported in the
transport of incoherent magnons in YIG films under microwave
\cite{nonlinear_microwave} or spin Hall effect \cite{not_drift,Luqiao_exp}
excitation. References~\cite{not_drift,Luqiao_exp} appear to support our
results without having to resort to spin-orbit interactions.

\begin{figure}[th]
{\includegraphics[width=8.6cm]{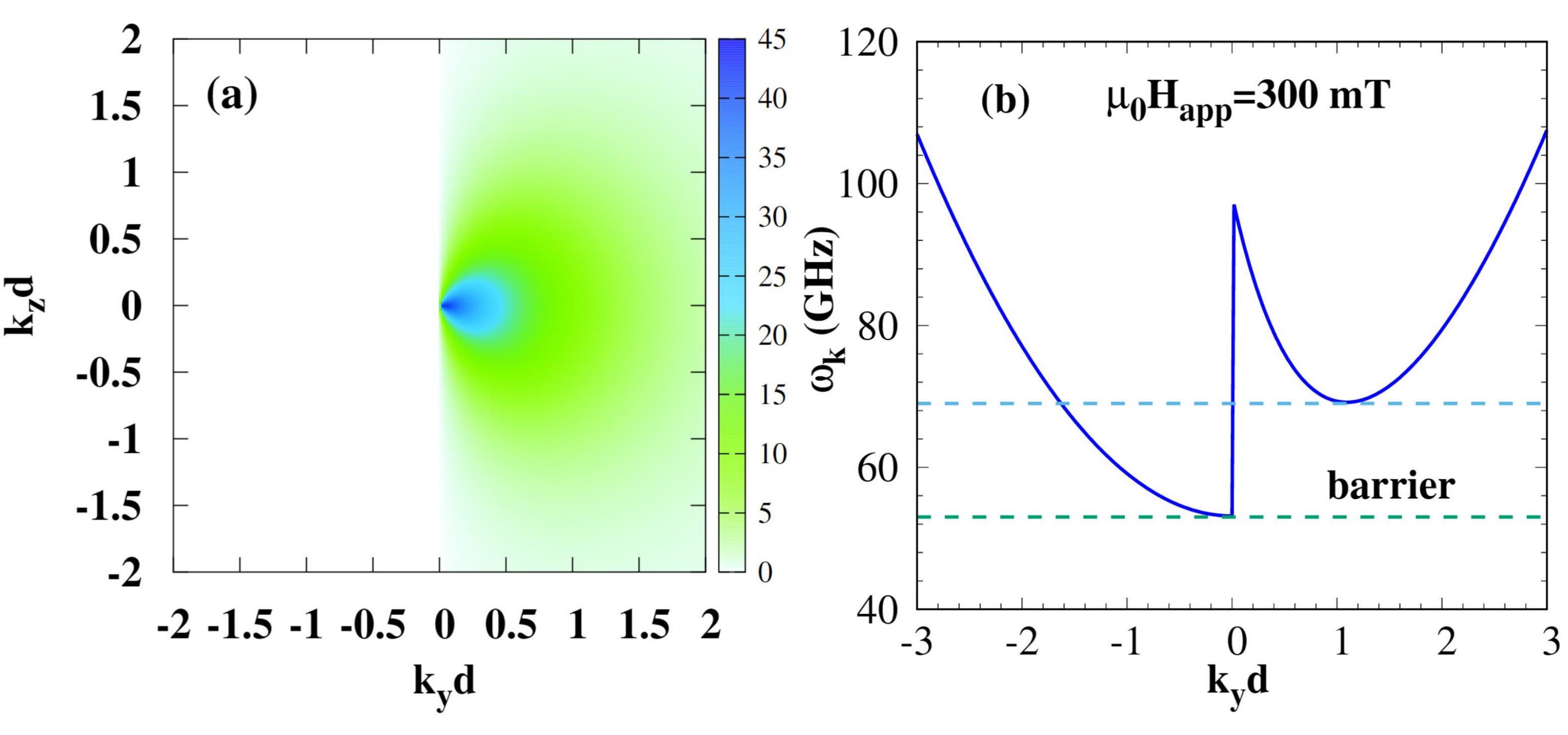}} \caption{Chirality of the
frequency shift [(a)] and spin wave dispersion with momentum $k_{y}%
\hat{\mathbf{y}}$ [(b)] induced by a superconducting gate. The frequency shift
for positive $k_{y}$ bars spin wave propagation in a large frequency window of
$\sim20~\mathrm{GHz}$ as indicated by the region between the dashed lines in
(b). A field $\mu_{0}H_{\mathrm{app}}=300$~mT is applied in the $\mathbf{\hat
{z}}$-direction.}%
\label{frequency_shift}%
\end{figure}

For a superconductor $\sigma_{c}(\omega)$ is complex and can be treated by a
two-fluid model
\cite{{superconductor_1,superconductor_2,superconductor_3,superconductor_4}}.
For simplicity we assume sufficiently low temperatures to freeze out the
quasi-particle excitations, but the effect sustains when $T\rightarrow0.85
T_{c}$ \cite{supplement}. The conductivity $\sigma_{c}(\tilde{\omega
}_{\mathbf{k}})=i{n_{s}e^{2}}/(m_{e}\tilde{\omega}_{\mathbf{k}})$ is then
purely imaginary, leading to a frequency shift%
\begin{equation}
\delta\omega_{\mathbf{k}}=\frac{d}{4}e^{-|\mathbf{k}|\left(  \frac{s}%
{2}+d\right)  }\frac{1-e^{-\left\vert \mathbf{k}\right\vert s}}{\left\vert
\mathbf{k}\right\vert }\mu_{0}^{2}\gamma M_{s}\frac{n_{s}e^{2}}{m_{e} }\left(
\frac{k_{y}}{\left\vert \mathbf{k}\right\vert }+\frac{|k_{y}|}{\left\vert
\mathbf{k}\right\vert }\right)  ,
\end{equation}
which is real and positive definite. Its chirality is complete since
$\delta\omega_{\mathbf{k}}$ vanishes for negative $k_{y}$ and arbitrary
$k_{z}$. It is typically quite large: For a $d=40$~nm NbN superconducting film
with electron density $n_{s}=10^{29}/\mathrm{m}^{3}$ \cite{carrier_density} on
top of a $s=20~\mathrm{nm}$ YIG film the shift amounts to $\delta\omega
=45$~GHz when $k_{y}=\left\vert \mathbf{k}\right\vert \ll\{1/d,1/s\}.$ This
value corresponds to the Zeeman energy of an applied magnetic field of
$255$~mT. Figure~\ref{frequency_shift} summarizes these features for the
chirality of frequency shift [(a)] and spin wave dispersion with momentum
$k_{y}\hat{\mathbf{y}}$ [(b)]. The singularity in Fig.~\ref{frequency_shift}%
(b) around $k_{y}=0$ has no physical consequences, since the group velocities
are positive (negative) for $k_{y}>0$ ($k_{y}<0$), but vanishes for Kittel
mode at $k_{y}=0$.

\textit{Unidirectional blocking of magnons.}---The above results imply
attractive functionalities created by superconductivity in wave magnonics. A
superconducting top gate on a magnetic film forms a nearly perfect switch. It
is opaque for ballistic spin waves launched, but becomes fully transmitting
simply by heating it above the critical superconducting temperature.

The superconducting gate couples most efficiently to spin waves that propagate
normal to the magnetization (Damon-Eshbach configuration) as in
Fig.~\ref{frequency_shift}(b). In order to assess the effect we consider a
wide superconducting gate at $w_{1}\leq y\leq w_{2}$ with width $w_{2}-w_{1}\gg\lambda\gg1/k_{y},$ where $\lambda$ is a magnon relaxation
length. It acts as a potential barrier that reflects magnons below an energy
threshold just as electric gates in field effect transistors, but now with a
$\mathbf{k}$-dependent barrier height.

\begin{figure}[th]
\hspace{-0.04cm}{\includegraphics[width=4.5cm,trim=0cm 0.7cm 0cm
0.7cm]{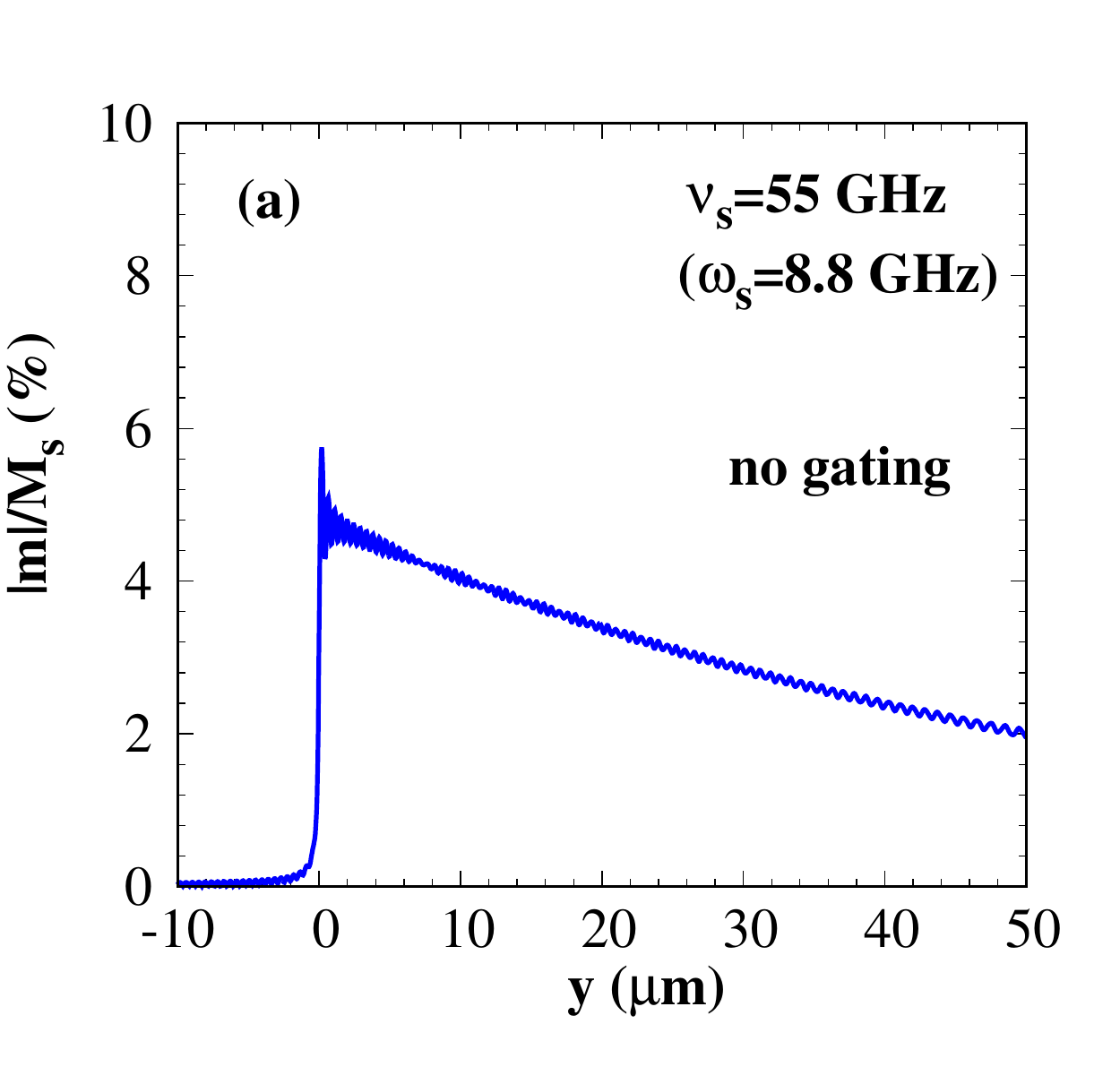}} \hspace{-0.41cm}%
{\includegraphics[width=4.5cm,trim=0cm 0.7cm 0cm
0.7cm]{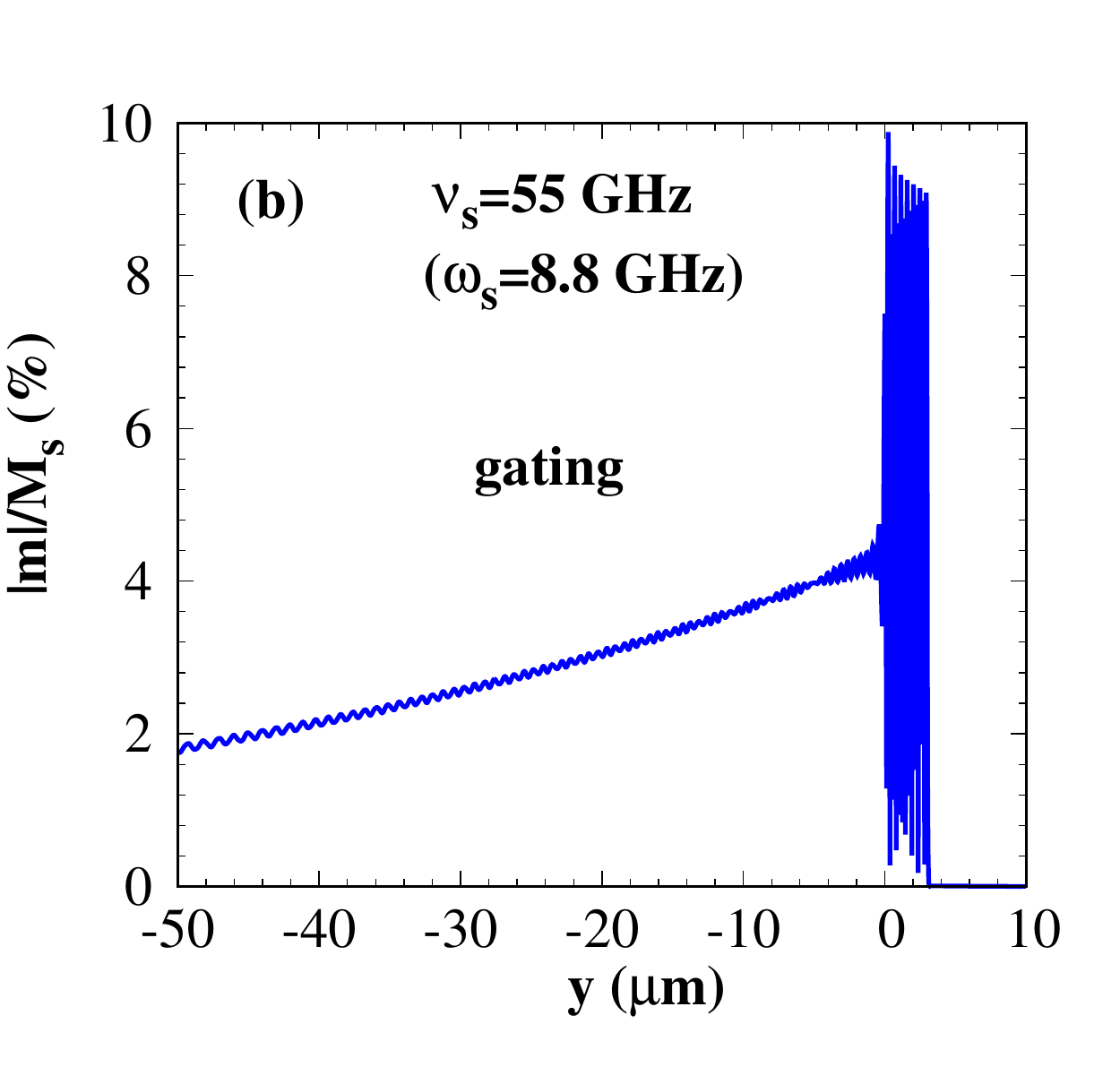}}\caption{Excited magnetization $|\mathbf{m}|$ by a
narrow stripline at $y=0$ without (a) and with (b) a superconducting gate that
covers the region from $3$ to $100~\mathrm{\mu m}$. The spin waves are totally
reflected at the gate edge and leak into the left half space.}%
\label{blocking}%
\end{figure}

We place a stripline with thickness $d_{s}=80$~nm and width $w_{s}=150$~nm
centered at the origin biased by a uniform ac current of frequency $\nu
_{s}=2\pi\times\omega_{s}=55$~GHz and current density $I_{s}=10^{6}%
~\mathrm{A/cm^{2}}$. When parallel to the external dc field and magnetization
in the underlying YIG, it\ launches spin waves with positive $k_{y}^{\ast
}=2\pi/0.45$~$\mathrm{\mu m}^{-1}$ into half space as shown in
Fig.~\ref{blocking}(a) in the form of a snapshot of the excited magnetization
$\left\vert \mathbf{m}\right\vert $
\cite{Chiral_pumping_Yu,Toeno_NV,stripline_poineering_1,stripline_poineering_2,Yu_review}%
. The Gilbert damping $\alpha_{G}=10^{-4}$ governs the decay length
$\lambda\sim\sqrt{(\alpha_{\mathrm{ex}}\mu_{0}\gamma M_{s})(\nu_{s}-\mu
_{0}\gamma H_{\mathrm{app}})}/(\alpha_{G}\nu_{s})=24.2~\mathrm{\mu m}$, but it
is not a simple exponential since the stripline of finite width also emits
waves around $k_{y}^{\ast}$ that cause the observed interference pattern. A
superconducting gate made from a NbN film with thickness $d=40$~nm covers the
YIG film from $y=w_{1}=3~\mathrm{\mu}$m to $w_{2}=100~\mathrm{\mu}$m. Below
the gate $w_{1}\leq y\leq w_{2}$, the diamagnetic field ($k_{y}^{\ast}%
d\ll1,k_{y}^{\ast}s\ll1$)
\begin{align}
\tilde{H}_{x}(y,t) &  =-\frac{n_{s}e^{2}}{m_{e}}\mu_{0}\frac{ds}{4}\left(
m_{x}(y,t)+\frac{1}{\omega_{s}}\frac{dm_{y}(y,t)}{dt}\right)  ,\nonumber\\
\tilde{H}_{y}(y,t) &  =-\frac{n_{s}e^{2}}{m_{e}}\mu_{0}\frac{ds}{4}\left(
-\frac{1}{\omega_{s}}\frac{dm_{x}(y,t)}{dt}+m_{y}(y,t)\right)  ,\nonumber
\end{align}
enters the LLG equation. We solve the time dependent problem in the steady
state and show representative $\left\vert \mathbf{m}\right\vert \equiv
\sqrt{m_{x}^{2}+m_{y}^{2}}$ in Fig.~\ref{blocking}(b). Technique details are
referred to the Supplemental Material \cite{supplement}. We clearly observe
the total reflection of spin waves at the gate edge due to the excited Oersted
field from the superconductor that only exists at its edge \cite{supplement}.
The reflected spin waves penetrate the left half-space in Fig.~\ref{blocking}%
(b), which remains silent in the absence of the gate [Fig.~\ref{blocking}(a)]
because the pumping is chiral. The amplitude of the magnetization between
source and gate $0<y<w_{1}$ is enhanced by a factor 2, i.e., incoming and
reflected spin waves interfere constructively. The reflection at the stripline
is very weak and cannot generate standing waves in the region $0<y<w_{1}$.
Nevertheless, a large number of coherent emitted and reflected waves with
different wave lengths coexist and cause complex interference fringes (refer
to Supplemental Material \cite{supplement} with other parameters). On the
other hand, the transmission of spin waves that impinge from the right of the
superconducting strip are not affected at all, i.e., the device acts as a
spin-wave isolator. Replacing the superconductor by $d=40$~nm copper strip, we
only enhance the damping of the spin waves without causing reflection
\cite{supplement}.

\textit{Spin-wave confinement.}---When we rotate the magnetization by 90$%
{{}^\circ}%
$ to become normal to the gate and stripline, the emitted spin waves
$\mathbf{k}=\left(  0,0,k_{z}\right)  $ propagate parallel to the
magnetization. The frequency shift of spin waves with positive and negative
$k_{z}$ under the gates on both sides of the stripeline as sketched in
Fig.~\ref{frequency_shift}(a) are smaller than Damon-Eshbach configuration,
but still very substantial. The device therefore forms a cavity, trapping the
spin waves analogous to, but without having to nanofabricate, a magnetic film.
The standing waves modulate the density of states that are strongly enhanced
at the subband edges, which may lead to strong coupling to the stripline
photons and other degrees of freedom.

\begin{figure}[th]
\hspace{-0.05cm}%
{\includegraphics[width=4.3cm,trim=0cm 0cm 0cm 0.2cm]{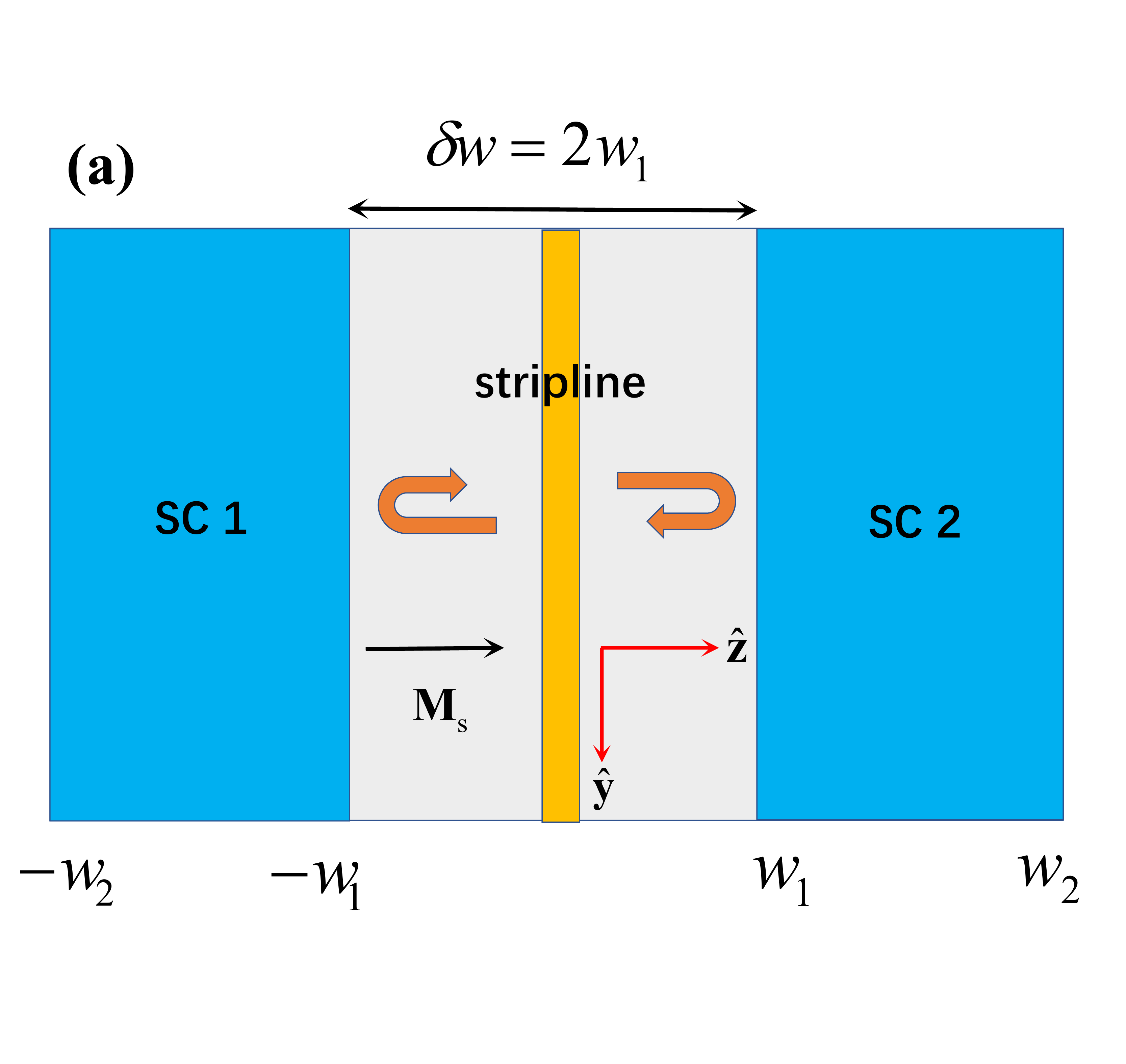}}
\hspace{-0.15cm}{\includegraphics[width=4.4cm,trim=0cm 0.3cm 0cm
0.5cm]{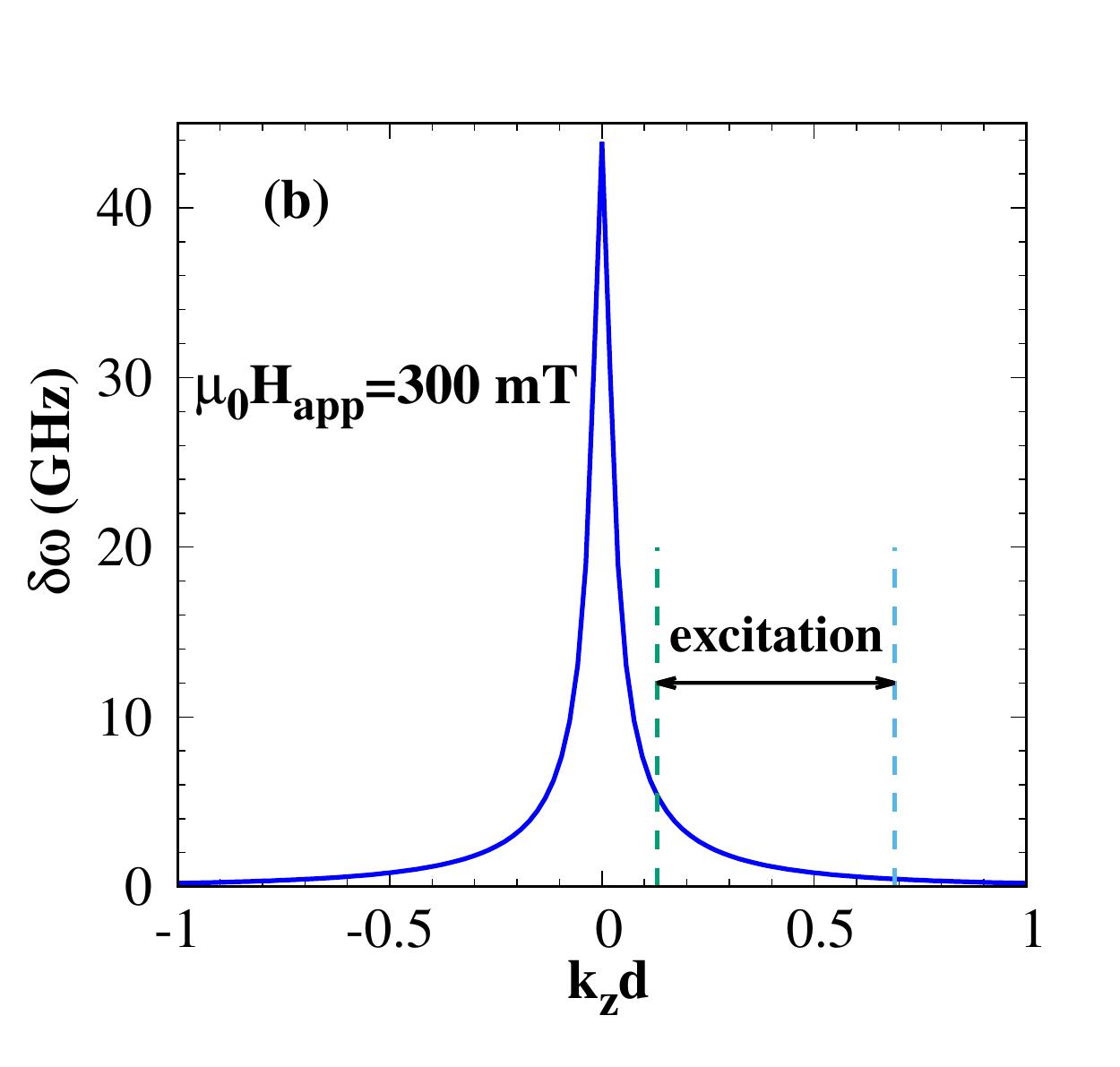}}\newline\hspace{-0.25cm}
{\includegraphics[width=4.45cm,trim=0.2cm 0.4cm 0.2cm 1.5cm]{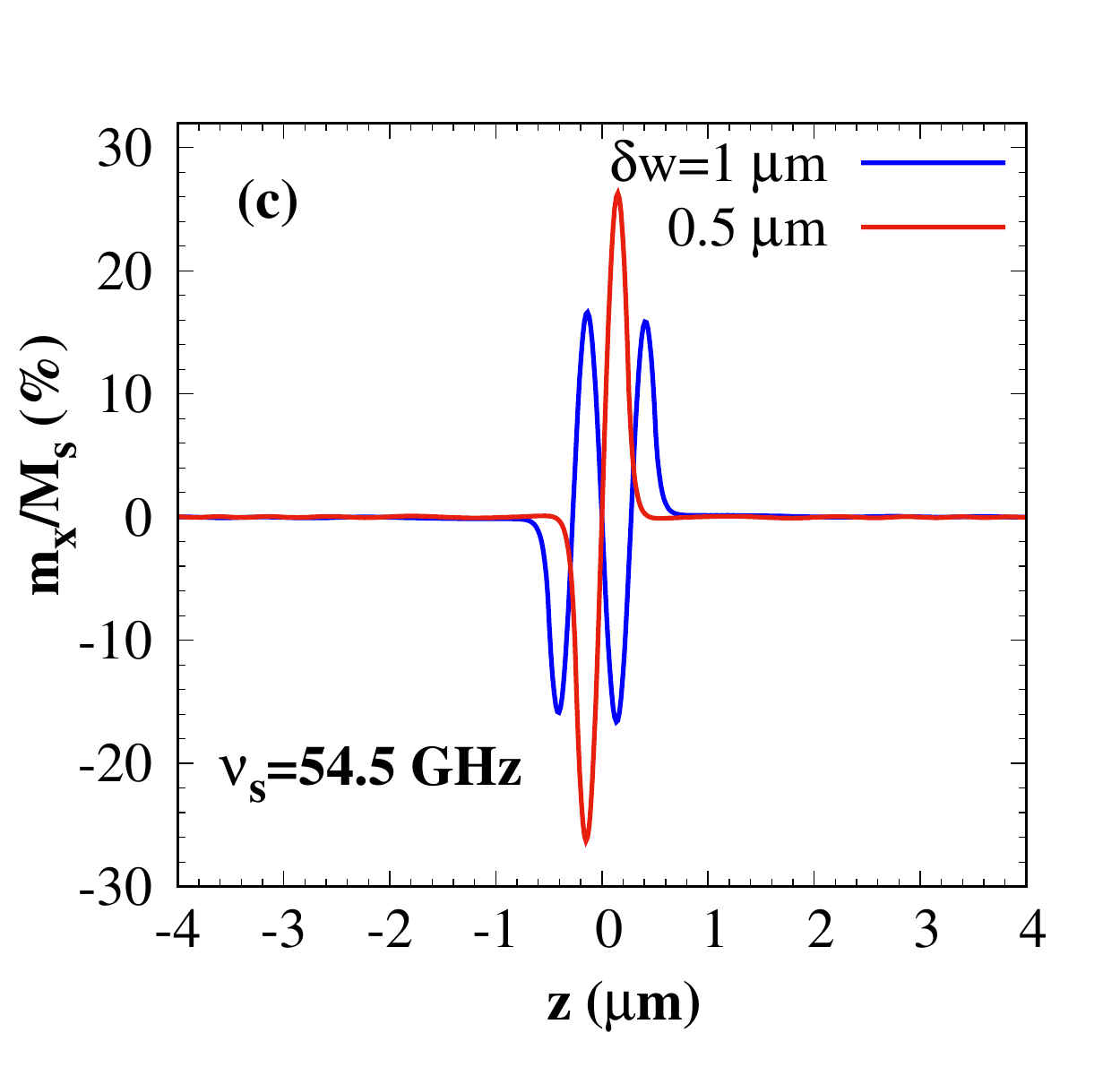}}
\hspace{-0.5cm} {\includegraphics[width=4.45cm,trim=0.2cm 0.4cm 0.2cm
1.5cm]{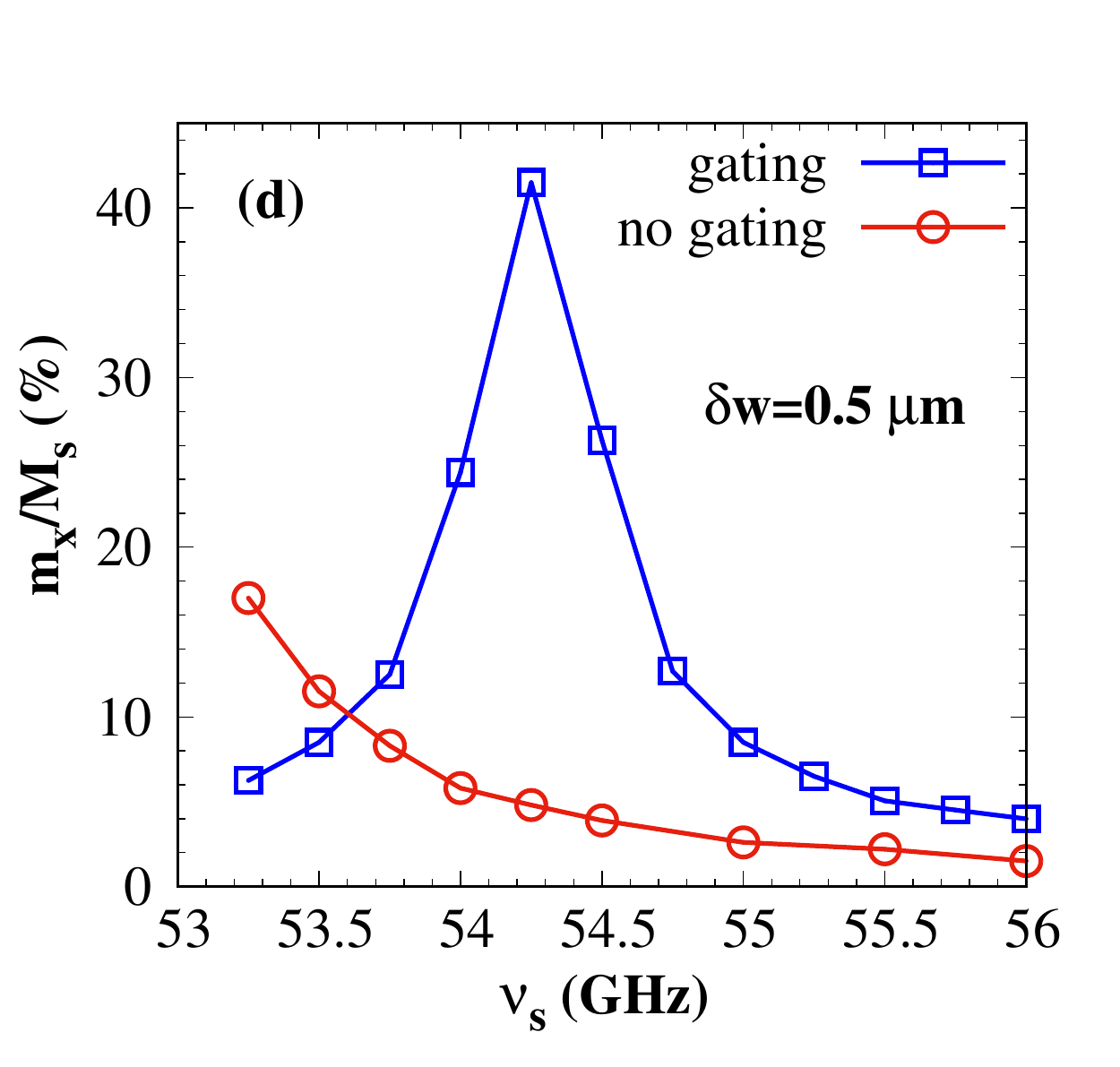}} \caption{Spin waves trapped by superconducting gates on both
sides of a microwave stripline atop a magnetic thin film magnetized along
$\hat{\mathbf{z}},$ \textit{i.e.}, normal to the gate. (a) is a top view of
the device. (b) shows the frequency shift of magnons under a uniform gate with
wave number $k_{z}$. The microwave frequencies in (c) and (d) lie in the
window indicated by the vertical dashed lines. In (c), we plot snapshots of
the maximum amplitudes $m_{x}$ for two gate distances $\delta
w=\{1,0.5\}\,\mathrm{\mu m}$ at a frequency close to the standing wave
frequencies with index $n=\{3,1\}$. In (d) we plot the excitation efficiencies
with and without gating for the $n=1$ resonance of the $0.5\,\mathrm{\mu m}$
cavity.}%
\label{cavity}%
\end{figure}

We substantiate these expectations by numerical modelling for a YIG\ film with
two superconducting gates located at $w_{1}\leq\left\vert z\right\vert \leq
w_{2}$ on both sides of a stripline at the origin as illustrated in
Fig.~\ref{cavity}(a). The spin waves now feel the back-action magnetic field
$\tilde{H}_{x}(z,t)=-({n_{s}e^{2}}/{m_{e}})\mu_{0}{ds}m_{x} (z,t)/4$ and
$\tilde{H}_{y}(z,t)=0$. We self-consistently solve the LLG equation
\cite{supplement} for cavity widths $\delta w=2w_{1}=1$ and $0.5\,\mathrm{\mu
m}$ and a stripline frequency interval $\nu_{s}\in\lbrack53.25,56]$~GHz
indicated in Fig.~\ref{cavity}(b), while the other parameters are the same as
in Fig.~\ref{blocking}. We choose frequencies typical in propagating magnon
spectroscopy and stripline widths that can be deposited by state-of-the-art
fabrication techniques. The stripline now excites spin waves with equal
amplitude to both sides that are reflected by the superconducting gates and
interfere. The steady states in Fig.~\ref{cavity}(c) are nearly perfectly
trapped standing spin waves with odd symmetry. The standing wave resonance for
a hard-wall potential are $\nu_{n}=\mu_{0}\gamma H_{\mathrm{app}}%
+\alpha_{\mathrm{ex}}\mu_{0}\gamma M_{s}(n\pi/\delta w)^{2}$, \textit{viz.}
$\nu_{1}\left(  \delta w=0.5~\mathrm{\mu m}\right)  =53.6$~GHz and $\nu
_{3}\left(  \delta w=1~\mathrm{\mu m}\right)  =54$~GHz. The boundary pinning
strongly suppresses the Kittel mode at $\nu_{0}=53.2$\thinspace GHz\textit{.}
The excitation frequency $\nu_{s}=54.5$~GHz is close to the resonances of both cavities.

Figure~\ref{cavity}(d) illustrates the enhanced excitation efficiency at a
standing wave resonance. The maximum at $\nu_{s}\sim54.25$~GHz is not much
above the hard-wall estimate of $53.6$~GHz, indicating efficient confinement.
However, the resonance widths of $\sim1$~GHz corresponds to a Gilbert damping
of $\alpha_{G}=0.02$ that is much larger than the intrinsic one, indicating
substantial gate leakage that can be suppressed by using a thicker
superconducting film. Nevertheless, comparison with the ungated result in
Fig.~\ref{cavity}(d) already shows a cavity enhancement of the magnetization
dynamics by an order of magnitude at the resonance.

Microstructuring of YIG films into stripes is usually accompanied by
substantial deterioration of the magnetic quality \cite{Qi_wang_strip}. The
trapping of spin waves by superconducting gates only requires deposition
of\ two metal films with a finite gap without introducing additional
roughness. Moreover, the option to modulate the trapping offers an easy
reprogramming of magnonic logical circuits.

\textit{Discussion and conclusion.}---The predicted total unidirectional
reflection is surprising and seems to violate thermodynamic principles.
However, we address here a coherent scattering process in a
finite system and not a diode-like transport between reservoirs.
Furthermore, the physics of chiral reflection of spin waves under a
superconducting gate is very different from conventional potential scattering.
The reflection of spin waves is a complicated process, in which an approaching
wave packet adiabatically generates the diamagnetic current in the
superconductor and the associated magnetic fields that push up the magnon gap.
The non-equilibrium scattering problem of coherent spin waves
injected by an external source that interact with a superconductor gate is not
subject to the constraints from linear response. 

In conclusion, we predict a non-reciprocal frequency shift of spin waves by
the diamagnetism of nearby superconductors, which leads to functionalities
such as unidirectional total reflection and confinement of spin waves in
magnetic films by top gates. When the gates turn into normal conductors the
chiral frequency shift turns into a chiral damping. Both effects enrich the
tool box of information communication and processing technology in photonics
\cite{nano_optics,chiral_optics}, plasmonics
\cite{plasmonics_1,plasmonics_2,Nori}, acoustics \cite{acoustic_1,acoustic_2},
electronics \cite{electronics_1,electronics_2}, superconductivity
\cite{superconductivity_1,superconductivity_2}, and spintronics
\cite{chiral_spintronics_1,chiral_spintronics_2}.

\vskip0.25cm \begin{acknowledgments}
This work is financially supported by the startup grant of
Huazhong University of Science and Technology (Grants No. 3004012185 and No. 3004012198) as well as JSPS KAKENHI Grant No. 19H00645.
\end{acknowledgments}

\end{document}